\documentclass[10pt,letterpaper]{article}
\usepackage{opex3}

\renewcommand{\hbar}{h^{\hspace{-2.2mm}-}}

\begin{document}

%%%%%%%%%%%%%%%%%% title page information %%%%%%%%%%%%%%%%%%
\title{Bends and splitters in graphene nanoribbon waveguides}

\author{Xiaolong Zhu, Wei Yan, N. Asger Mortensen, and Sanshui Xiao}

\address{DTU Fotonik - Department of Photonics Engineering, \\Technical
University of Denmark, DK-2800 Kongens Lyngby, Denmark,\\
and Center for Nanostructured Graphene (CNG), \\
Technical University of Denmark, DK-2800 Kgs. Lyngby, Denmark}

\email{saxi@fotonik.dtu.dk} %% email address is required


\begin{abstract}
We investigate the performance of bends and splitters in graphene nanoribbon waveguides.
Although the graphene waveguides are lossy themselves, we show that bends and splitters do not
induce any additional loss
provided that the nanoribbon width is sub-wavelength. We use
transmission line theory to qualitatively interpret the behavior observed in our simulation.
Our results pave a promising way to realize ultra-compact devices operating in the terahertz region.
\end{abstract}

\ocis{(230.7370) Waveguides; (240.6690) Surface waves; (130.3120) Integrated optics devices.} % REPLACE WITH CORRECT OCIS CODES FOR YOUR ARTICLE

%%%%%%%%%%%%%%%%%%%%%%%% References %%%%%%%%%%%%%%%%%%%%%%%%%
%\bibliographystyle{osajnl}
%\bibliography{References_New}

\section{Introduction}
Graphene has attracted considerable attention
due to its unique electronic and optical properties~\cite{Novoselov2004P1,Novoselov2005P1,Vakil2011P1,Grigorenko2012P1,Bao2012P1}.
The high carrier mobility approaching 200,000 cm$^2$V$^{-1}$s$^{-1}$ offers the possibility of building
high-speed optoelectronic devices, such as photodetectors~\cite{Xia2009P1} and optical modulators~\cite{Liu2011P1}.
Despite being a single atomic layer, graphene can couple strongly to light, enabling a constant
absorbance of $\pi a=2.3\%$ in the infrared-to-visible spectral range~\cite{Mak2008P1}, where $a=e^2/\hbar c$ is the
fine-structure constant. More importantly, the optical properties of graphene can be substantially modified by changing the Fermi energy $E_F$~\cite{Li2008P1,Wang2008P1}.
For a slightly doped graphene ($E_F<\hbar\omega/2$), the conductivity of graphene is dominated by the interband contributions. In the terahertz range when $E_F>\hbar\omega/2$, the intraband contribution becomes relevant, thus allowing graphene to support graphene-plasmon polaritons (GPPs), i.e., the collective oscillation of carriers. Compared to conventional plasmonic materials, e.g. gold, doped-graphene plasmon has promising properties including tunability, extreme field confinement and low propagation loss.

Graphene-plasmon polaritons supported by doped graphene have been widely investigated~\cite{Jablan2009P1,Thongrattanasiri2012P1,Gao2012P1,Ju2011P1,Yan2012P1}.
It has been shown that remarkable optical absorption enhancement~\cite{Thongrattanasiri2012P1,Gao2012P1,Zhan2012P1} in the graphene can be achieved by patterning the single sheet of doped graphene. Moreover, the GPP can also be confined laterally, e.g., in graphene ribbons.
Recently, much attention has been focused on investigating propagation properties of GPP in graphene nanoribbons~\cite{Nikitin2011P1,Christensen2012P1}, which can be used as a promising candidate to construct future high-speed compact-devices working in the terahertz region.
Obviously it is indispensable to have graphene nanoribbon bends and splitters as the basic structures in these compact devices. However, there is a lack of characterization of the GPP in these geometries. In this paper, we investigate the performance of bends and power splitters in graphene nanoribbon waveguides, particularly focusing on the question whether these bends and splitters will induce reflection or excess scattering loss on top of the propagation loss in the nanoribbon waveguides themselves. Our analysis in this paper relies on a classical electromagnetic description where graphene is described by a local conductivity, and the results presented here do not include the microscopic details of the edges of graphene nanoribbons~\cite{Nikitin2011P1}.
\begin{figure}[htbp]
\centering\includegraphics[width=13 cm]{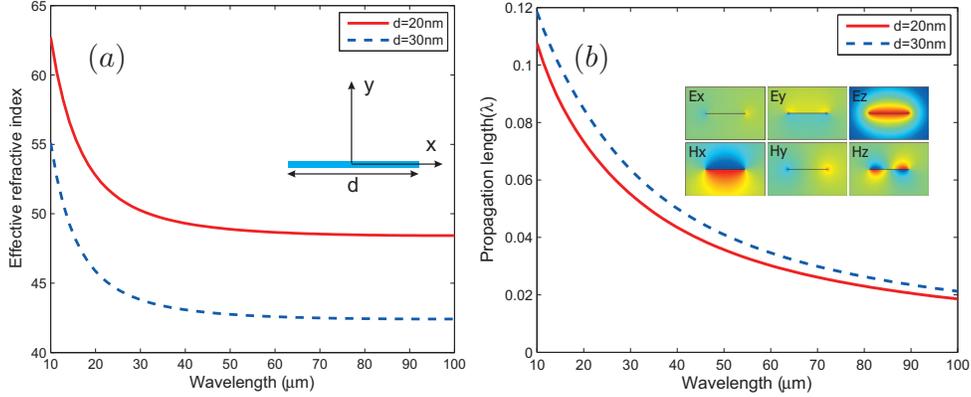}
\caption{(a) Effective refractive indices of GPP modes as a function of wavelength for freestanding graphene ribbons of 20~nm (the solid line) and 30~nm (the dashed line) widths. Inset illustrates a cross-section graphene nanoribbon waveguide with the width d, where the axis indicates light propagation along the $z$ direction. (b) Propagation lengths as a function of wavelength for the freestanding graphene ribbons. Inset shows field distributions of the GPP mode at $\lambda=15~\mu$m for the case of d~$=$~30~nm. }
\label{dispersion}
\end{figure}
\section{Mode properties of GPP in a graphene nanoribbon}
Let us first investigate mode properties of GPP in a freestanding graphene nanoribbon, as shown in the inset of Fig.\ref{dispersion} (a), with an ultra-thin width d$<$50~nm. The inset in Fig.~\ref{dispersion}(a) illustrates a cross-section of the nanoribbon and its corresponding axis.
Due to the translational symmetry in the $z$ direction, the electric solution of the nanoribbon waveguide has the form
$\vec{\textbf{E}}(\vec{r},\omega,t)=\vec{\textbf{E}}(x,y)exp(ik_z(\omega) z-i\omega t)$, where $k_z(\omega)$ is the wave vector in the propagation direction $z$.
For a waveguide in which the material is lossy, $k_z(\omega)$ is complex. An effective refractive index $n_{\textrm{\tiny {eff}}}$ of the waveguide mode is defined by $n_{\textrm{\tiny {eff}}}=Re(k_z(\omega))/k_0$, and the propagation length $L$ of the mode is given by $L=1/[2 Im(k_z(\omega))]$, where $k_0=2\pi/\lambda_0$ (wavelength in vacuum $\lambda_0$).
In the simulations, the graphene ribbon is modeled as a thin layer of thickness t~$=$~0.5~nm and characterized
by a dielectric function $1+i\sigma/(\omega t\epsilon_0)$, where $\sigma$ is the surface conductivity of graphene,
$\epsilon_0$ is the vacuum permittivity, and $\omega$ is the light frequency. Note that the simulation results do
not rely on the thickness of the film as long as the meshing is sufficiently fine.
In this paper we consider the case where the photon energy is always less than $2E_F$,
and under this circumstance the intraband contribution to the conductivity dominates. The intraband conductivity
$\sigma_{\textrm{\tiny {intra}}}$ of the graphene film can be taken from the standard random-phase
approximation~\cite{Hwang2007P1,Falkovsky2007P1}
as follows
\begin{eqnarray}
\sigma_{\textrm{\tiny {intra}}}(\omega)=\frac{ca}{\pi}\frac{iE_F}{\hbar\omega+i\hbar/\tau}\left(\frac{2k_BT}{E_F}\right)\ln\left[2\cosh\left(\frac{E_F}{2k_BT}\right)\right],
\label{sigma}
\end{eqnarray}
where $k_BT$ is the thermal energy, and $\tau$ is the relaxation time.
$\tau$ can be evaluated by $\tau=\mu E_F/e\upsilon_F^2$,
where $\upsilon_F\approx c/300$ is the Fermi velocity and $\mu=10,000$~cm$^2$V$^{-1}$s$^{-1}$
is a typically measured, impurity-limited DC mobility~\cite{Novoselov2004P1,Novoselov2005P1}.
Then $\tau$ is approximately $3\times10^{-13}$~s for $E_F=0.3$~eV. In this work, we consider Fermi energy of 0.3~eV for all calculations,
and the mode calculations have been performed by using a commercial available finite-element-method software package (COMSOL).

Figure~\ref{dispersion} (a) shows the effective refractive indices $n_{\textrm{\tiny {eff}}}$ of the GPP supported by freestanding nanoribbons for two different widths (d~$=$~20,~30~nm). In the frequency range of our interest, the nanoribbons with the widths of $20$~nm and $30$~nm only support the fundamental waveguide mode. It has been shown that a wide graphene ribbon may support many modes including edge modes and "bulk" modes~\cite{Nikitin2011P1,Christensen2012P1}. With the decrease of the ribbon width, there exists cut-offs for the high-order modes. Eventually, the nanoribbon only has the fundamental mode with even symmetry, similar to the dispersion of the surface-plasmon in a metal strip~\cite{Maierbook}. The effective refractive indices shown in Fig.~\ref{dispersion} (a) of the fundamental GPP of the nanoribbons have large values, e.g. 48 for the case of $\lambda_0=15~\mu$m, and d~$=$~30~nm, indicating that the GPP mode is extremely confined near the nanoribbon. This is also confirmed by the mode profile shown in the inset of Fig.~\ref{dispersion} (b). The effective refractive index becomes larger when decreasing the nanoribbon width from 30~nm to 20~nm. Compared to the nobel metals, graphene generally has a relatively large conductivity, thus leading to a lower propagation loss. Propagation lengths $L$ as a function of the wavelength are plotted in Fig.~\ref{dispersion} (b).
As an example, when $\lambda_0=10~\mu$m, and d~$=$~30~nm, $L$ reaches $0.12\lambda_0$, equivalent to 1.2$~\mu$m. The propagation loss of the GPP becomes weaker when the wavelength decreases, which is mainly attributed to the optical property of the graphene material.
On the other hand, the propagation length becomes shorter by shrinking the nanoribbon width for a specific working wavelength.
It can be well understood by the fact that when the effective refractive index of the GPP mode becomes larger, there will be a larger field overlap with the graphene, thus leading to a higher propagation loss. Similar to surface plasmons supported by a metallic waveguide, there is also a competition in the graphene ribbon waveguide between the mode confinement and propagation length~~\cite{Maier2003P1,Bozhevolnyi2006P1,Pile2004P1,Xiao2010P2}.
\begin{figure}[htbp]
\centering\includegraphics[width=6 cm]{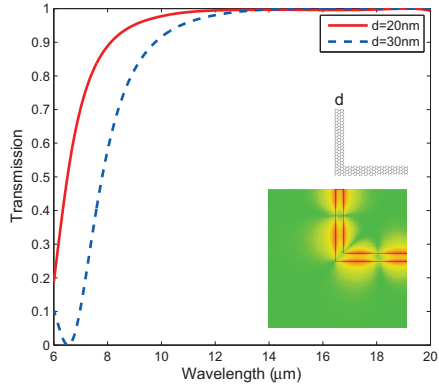}
\caption{Transmission spectra of graphene nanoribbon waveguide bends for d~$=$~20, 30~nm. Inset illustrates its top-view corresponding structure and the top-view magnetic amplitude at $y=0$ when $\lambda_0=15~\mu$m and d~$=$~30~nm.}
\label{bending}
\end{figure}
\section{Bends in graphene nanoribbon waveguides}
We now turn to the question whether waveguide bends (breaking translational invariance) lead to reflection or excess absorption? As mentioned above, the graphene nanoribbon waveguide is generally lossy. In order to evaluate the performance of the bends, we calculate the transmission coefficient and normalize it to that of a straight graphene ribbon waveguide with the same propagation length. Propagation properties of the graphene nanoribbon waveguide are analyzed by
a finite-integral-method software package (CST microwave studio).
In all cases, the width d of the graphene nanoribbons is chosen to be much smaller than the wavelength, so that only the fundamental waveguide mode exists, see the discussion above.

Figure~\ref{bending} presents the normalized transmission through a $90^o$ sharp graphene nanoribbon bend, therefore the additional loss associated with reflection and excess scattering loss for the sharp graphene waveguide can be obtained. One can see in Fig.~\ref{bending} that there is no additional loss at longer wavelengths. Note that a similar high transmission through sharp bends can also be realized in the photonic crystal systems~\cite{Mekis1996P1,Joannopoulos,Xiao2005P1}.
For the case of d~$=$~30~nm, the limiting wavelength at which the transmission coefficient
decreases below 99$\%$ is 14~$\mu$m. The limiting wavelength is blue-shifted to 12~$\mu$m when decreasing the nanoribbon width from 30~nm to 20~nm.
To evaluate the transmittance through the sharp graphene ribbon, we can also employ the analogy between the subwavelength waveguiding in photonics and electronics and use a common quasistatic approximation~\cite{Veronis2005P1,Matsuzaki2008P1}.
With this analogy and quasistatic approximation (valid as long as d~$\ll\lambda$), we can consider the sharp bend equivalent to a junction between two transmission lines with the same characteristic impedance. Within the quasistatic approximation no additional loss emerges for the bend. It is naturally understood that the operating wavelength range for having additional bend loss widens with the decrease of d. The inset in Fig.~\ref{bending} shows the top-view magnetic field amplitude distribution at $y=0$ when light passes through the sharp corner for the case of d~$=$~30~nm at $\lambda_0=15~\mu$m. To illustrate this behavior clearly, we neglect the material loss of the graphene. One can see that light can be perfectly turned by 90$^o$ without any noticeable back-scattering or radiation.
\begin{figure}[htbp]
\centering\includegraphics[width=6 cm]{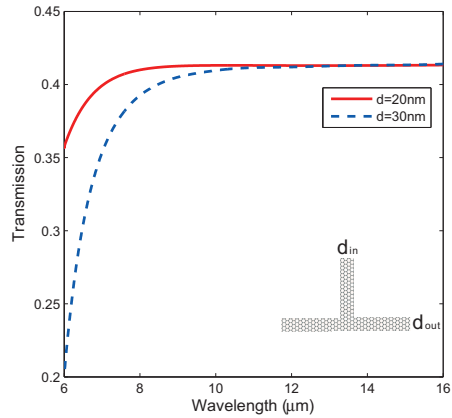}
\caption{Transmission spectra of graphene nanoribbon T-shape splitters (shown in the inset with the top view) for d~$=$~20, 30~nm. In this case we choose d$_{\textrm{\tiny {in}}}$~$=$~d$_{\textrm{\tiny {out}}}$~$=$~d.}
\label{splitter}
\end{figure}

\section{Splitters in graphene nanoribbon waveguides}
\begin{figure}[htbp]
\centering\includegraphics[width=7 cm]{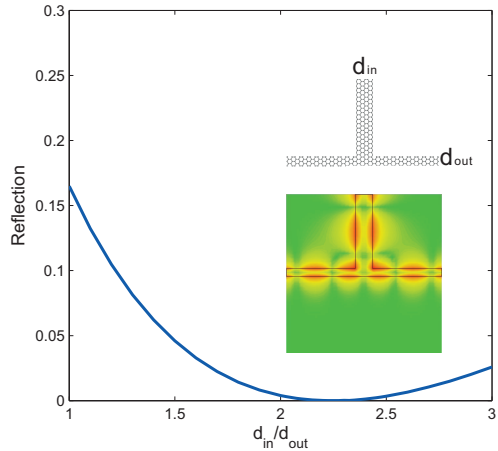}
\caption{Reflection coefficient of a graphene nanoribbon T-shape splitter as a function of $\textrm{d}_\textrm{\tiny {in}}/\textrm{d}_\textrm{\tiny {out}}$ at $\lambda=12~\mu$m. Inset shows its top-view corresponding structure and illustrates zero-reflection when $\textrm{d}_\textrm{\tiny {in}}/\textrm{d}_\textrm{\tiny {out}}=2.25$ at $\lambda=12~\mu$m.}
\label{splitter2}
\end{figure}

Next, we focus on the propagation property of graphene nanoribbon splitters, which is another key element. The calculated normalized transmission coefficients as a function of wavelength for graphene T-shape nanoribbon waveguides
are plotted in Fig.~\ref{splitter} for the cases of d~$=$~20, 30~nm. The inset shows a top-view of the structure.
Similar to the result for the nanoribbon bend, the transmission at the output graphene waveguides becomes larger as the wavelength increases, and it saturates at the longer wavelengths, reaching ~42\%. We find that the limiting wavelengths are almost the same as those for the graphene bends shown in Fig.~\ref{bending}. Note that the widths for the input and output graphene waveguides have the same values as d. Due to the symmetry of the structure, the transmitted power is equally distributed between the two output waveguide branches, so that the total transmission is close to 84\%.
With the quasistatic approximation, we can replace the splitter configuration by an equivalent network consisting of a junction of three transmission lines with the same characteristic impedance $Z_0$. The combination of the two output transmission lines are connected to the input transmission line, thus resulting in the equivalent load impedance with a value of $Z_L$~$=$~2$Z_0$. The transmission coefficient at each output governed by $1-[(Z_L-Z_0)/(Z_L+Z_0)]^2$ is close to 88\%. The result obtained by the transmission lines has a small deviation when compared with our simulated result.
When using the transmission line theory, one has to assume that the energy of the guided mode is mainly confined in the waveguide and that the distribution of the transverse electromagnetic field is nearly uniform within the waveguide. This is only qualitatively correct for the GPP supported by the graphene nanoribbon waveguide. However, we can still use the transmission line theory to qualitatively interpret the behavior observed in our simulation. In order to improve the transmission coefficient of the graphene nanoribbon splitter, we can adjust the characteristic impedance of the input waveguide $Z_{\tiny {in}}$ by changing the width d$_{\textrm{\tiny {in}}}$ of the input waveguide. Figure~\ref{splitter2} illustrates the calculated reflection coefficient of the
graphene T-shaped splitter at $\lambda_0=12~\mu$m as a function of d$_{\textrm{\tiny {in}}}$/d$_{\textrm{\tiny {out}}}$, where d$_{\textrm{\tiny {out}}}$, the thickness of the output waveguide branches, is fixed at 20~nm.
An optimal structure with nearly-zero reflection coefficient is observed at d$_{\textrm{\tiny {in}}}$/d$_{\textrm{\tiny {out}}}$~$=$~2.25.
The lower-inset in Fig.~\ref{splitter2} shows the top-view magnetic field amplitude distribution at $y=0$ when light passes through the T-shape configuration for the case of $\textrm{d}_\textrm{\tiny {in}}/\textrm{d}_\textrm{\tiny {out}}=2.25$ at $\lambda_0=12~\mu$m.

\section{Summary and discussion}
We have investigated the performance of bends and splitters in graphene nanoribbon waveguides, where the nanoribbon
width is much smaller than the working wavelength. We have shown that sharp bends and splitters with no additional
loss can be designed for longer wavelengths. The effective characteristic impendence model has been used to qualitatively
interpret the simulation results.
As discussed above, the GPP in the graphene nanoribbon has an extreme confinement, therefore making it difficult to
couple light into the nanoribbon by use of the butt-coupling technology. Similar to the surface plasmon supported by the
noble metals, a grating coupler could be used to excite the GPP in the nanoribbons. Recently, the propagating
and localized graphene-plasmon polaritons have been visualized
in real space by scattering-type scanning near-field optical microscopy~\cite{Chen2012P1,Fei2012P1}.
Our results further advance the potential of such devices by providing building blocks to construct graphene circuits for future ultra-compact terahertz devices.

\section*{Acknowledgments}
This work is partly supported by the Center for Nanostructured Graphene, sponsored by the Danish
National Research Foundation.


\begin{thebibliography}{10}
\newcommand{\enquote}[1]{``#1''}

\bibitem{Novoselov2004P1}
K.~S. Novoselov, A.~K. Geim, S.~V. Morozov, D.~Jiang, Y.~Zhang, S.~V. Dubonos,
  I.~V. Grigorieva, and A.~A. Firsov, \enquote{Electric field effect in
  atomically thin carbon films,} Science \textbf{306}, 666--669 (2004).

\bibitem{Novoselov2005P1}
K.~S. Novoselov, A.~K. Geim, S.~V. Morozov, D.~Jiang, M.~I. Katsnelson, I.~V.
  Grigorieva, S.~V. Dubonos, and A.~A. Firsov, \enquote{Two-dimensional gas of
  massless dirac fermions in graphene,} Nature \textbf{438}, 197--200 (2005).

\bibitem{Vakil2011P1}
A.~Vakil and N.~Engheta, \enquote{Transformation optics using graphene,}
  Science \textbf{332}, 1291--1294 (2011).

\bibitem{Grigorenko2012P1}
A.~N. Grigorenko, M.~Polini, and K.~S. Novoselov, \enquote{Graphene
  plasmonics,} Nat. Photonics \textbf{6}, 749--758 (2012).

\bibitem{Bao2012P1}
Q.~Bao and K.~P. Loh, \enquote{Graphene photonics, plasmonics, and broadband
  optoelectronic devices,} ACS Nano \textbf{6}, 3677--3694 (2012).

\bibitem{Xia2009P1}
F.~Xia, T.~Mueller, Y.~Lin, A.~Valdes-Garcia, and P.~Avouris,
  \enquote{Ultrafast graphene photodetector,} Nat. Nanotech. \textbf{4},
  839--843 (2009).

\bibitem{Liu2011P1}
M.~Liu, X.~Yin, E.~Ulin-Avila, B.~Geng, T.~Zentgraf, L.~Ju, F.~Wang, and
  X.~Zhang, \enquote{A graphene-based broadband optical modulator,} Nature
  \textbf{474}, 64--67 (2011).

\bibitem{Mak2008P1}
K.~F. Mak, M.~Y. Sfeir, Y.~Wu, C.~Lui, J.~A. Misewich, and T.~F. Heinz,
  \enquote{Measurement of the optical conductivity of graphene,} Phys. Rev.
  Lett. \textbf{101}, 196405 (2008).

\bibitem{Li2008P1}
Z.~Q. Li, E.~A. Henriksen, Z.~Jiang, Z.~Hao, M.~C. Martin, P.~Kim, H.~L.
  Stormer, and D.~N. Basov, \enquote{Dirac charge dynamics in graphene by
  infrared spectroscopy,} Nat. Phys. \textbf{4}, 532--535 (2008).

\bibitem{Wang2008P1}
F.~Wang, Y.~Zhang, C.~Tian, C.~Girit, A.~Zettl, M.~Crommie, and Y.~R. Shen,
  \enquote{Gate-variable optical transitions in graphene,} Science
  \textbf{206}, 206--209 (2008).

\bibitem{Jablan2009P1}
M.~Jablan, H.~Buljan, and Soljacic, \enquote{Plasmonics in graphene at infrared
  frequencies,} Phys. Rev. B \textbf{80}, 245435 (2009).

\bibitem{Thongrattanasiri2012P1}
S.~Thongrattanasiri, F.~H.~L. Koppens, and F.~J. Garcia~de Abajo,
  \enquote{Complete optical absorption in periodically patterned graphene,}
  Phys. Rev. Lett. \textbf{108}, 047401 (2012).

\bibitem{Gao2012P1}
W.~Gao, J.~Shu, C.~Qiu, and Q.~Xu, \enquote{Excitation of plasmonic waves in
  graphene by guided-mode resonances,} ACS Nano \textbf{6}, 7806--7813 (2012).

\bibitem{Ju2011P1}
L.~Ju, B.~Geng, J.~Horng, C.~Girit, M.~Martin, Z.~Hao, H.~A. Bechtel, X.~Liang,
  A.~Zettl, Y.~R. Shen, and F.~Wang, \enquote{Graphene plasmonics for tunable
  terahertz metamaterials,} Nat. Nanotech. \textbf{6}, 630--634 (2011).

\bibitem{Yan2012P1}
F.~Yang, J.~R. Sambles, and G.~W. Bradberry, \enquote{Tunable infrared
  plasmonic devices using graphene/insulator stacks,} Nature Nanotech.
  \textbf{7}, 330--334 (2012).

\bibitem{Zhan2012P1}
T.~R. Zhan, F.~Y. Zhao, X.~H. Hu, X.~H. Liu, and J.~Zi, \enquote{Band structure
  of plasmons and optical absorption enhancement in graphene on subwavelength
  dielectric gratings at infrared frequencies,} Phys. Rev. B \textbf{86},
  165416 (2012).

\bibitem{Nikitin2011P1}
A.~Y. Nikitin, F.~Guinea, F.~Garcia-Vidal, and L.~Martin-Moreno, \enquote{Edge
  and waveguide terahertz surface plasmon modes in graphene microribbons,}
  Phys. Rev. B \textbf{84}, 161407 (R) (2011).

\bibitem{Christensen2012P1}
J.~Christensen, A.~Manjavacas, S.~Thongrattanasiri, F.~Koppens, and F.~J.
  Garcia~de Abajo, \enquote{Graphene plasmon waveguiding and hybridization in
  individual and paired nanoribbons,} ACS Nano \textbf{6}, 431--440 (2012).

\bibitem{Hwang2007P1}
E.~H. Hwand and S.~Das~Sarma, \enquote{Dielectric function, screening, and
  plasmons in two-dimensional graphene,} Phys. Rev. B \textbf{75}, 205418
  (2007).

\bibitem{Falkovsky2007P1}
L.~A. Falkovsky and A.~A. Varlamov, \enquote{Space-time dispersion of graphene
  conductivity,} Eur. Phys. J. B \textbf{56}, 281--284 (2007).

\bibitem{Maierbook}
S.~A. Maier, \emph{Plasmonics: Fundamentals and Applications} (Springer, 2007),
  1st ed.

\bibitem{Maier2003P1}
S.~A. Maier, P.~G. Kik, H.~A. Atwater, S.~Meltzer, E.~Harel, B.~E. Koel, and
  A.~A.~G. Requicha, \enquote{Local detection of electromagnetic energy
  transport below the diffraction limit in metal nanoparticle plasmon
  waveguides,} Nat. Mater. \textbf{2}, 229--232 (2003).

\bibitem{Bozhevolnyi2006P1}
S.~I. Bozhevolnyi, V.~S. Volkov, E.~Devaux, J.-Y. Laluet, and T.~W. Ebbesen,
  \enquote{Channel plasmon subwavelength waveguide components including
  interferometers and ring resonators,} Nature \textbf{440}, 508--511 (2006).

\bibitem{Pile2004P1}
D.~F.~P. Pile and D.~K. Gramotnev, \enquote{Channel plasmon-polariton in a
  triangular groove on a metal surface,} Opt. Lett. \textbf{29}, 1069--1071
  (2004).

\bibitem{Xiao2010P2}
S.~Xiao, J.~Zhang, L.~Peng, C.~Jeppesen, R.~Malureanu, A.~Kristensen, and N.~A.
  Mortensen, \enquote{Nealy-zero transmission through periodically modulated
  ultrathin metal films,} Appl. Phys. Lett. \textbf{97}, 071116 (2010).

\bibitem{Mekis1996P1}
A.~Mekis, J.~C. Chen, I.~Kurland, S.~Fan, P.~R. Villeneuve, and J.~D.
  Joannopoulos, \enquote{High transmission through sharp bends in photonic
  crystal waveguides,} Phys. Rev. Lett. \textbf{77}, 3787--3790 (1996).

\bibitem{Joannopoulos}
J.~D. Joannopoulos, R.~D. Meade, and J.~Winn, \emph{Photonic Crystals: Modling
  the Flow of Light} (Princeton Univ. Press, Princeton, NJ, 1995), 1st ed.

\bibitem{Xiao2005P1}
S.~S. Xiao and M.~Qiu, \enquote{Study of transmission properties for waveguide
  bends by use of a circular photonic crystal,} Phys. Lett. A \textbf{340},
  474--479 (2005).

\bibitem{Veronis2005P1}
G.~Veronis and S.~H. Fan, \enquote{Bends and splitters in
  metal-dielectric-metal subwavelength plasmonic waveguides,} Appl. Phys. Lett.
  \textbf{87}, 131102 (2005).

\bibitem{Matsuzaki2008P1}
Y.~Matsuzaki, T.~Okamoto, M.~Haraguchi, M.~Fukui, and M.~Nakagaki,
  \enquote{Characteristics of gap plasmon waveguide with stub structures,} Opt.
  Express \textbf{16}, 16314--16325 (2008).

\bibitem{Chen2012P1}
J.~Chen, M.~Badioli, P.~Alonso-Gonzalez, S.~Thongrattanasiri, F.~Huth,
  J.~Osmond, M.~Spasenovic, A.~Centeno, A.~Pesquera, P.~Godignon, A.~Z. Elorza,
  N.~Camara, F.~J. Garcia~de Abajo, R.~Hillenbrand, and F.~H.~L. Koppens,
  \enquote{Optical nano-imaging of gate-tunable graphene plasmons,} Nature
  \textbf{487}, 77--81 (2012).

\bibitem{Fei2012P1}
Z.~Fei, A.~S. Rodin, G.~O. Andreev, W.~Bao, A.~S. McLeod, M.~Wagner, L.~M.
  Zhang, Z.~Zhao, G.~Thiemens, M.~andDominguez, M.~M. Fogler, A.~H.
  Castro~Neto, C.~N. Lau, K.~F., and D.~N. Basov, \enquote{Gate-tuning of
  graphene plasmons revealed by infrared nano-imaging,} Nature \textbf{487},
  82--85 (2012).

\end{thebibliography}
\end{document}